\documentclass[11pt, preprint]{aastex631} 

\usepackage{bm}          
\usepackage{graphicx}    
\usepackage{amsmath}     
\usepackage{natbib}      
\usepackage{aas_macros}  

\newcommand{\mum}{\thinspace\rm{\mu m}}
\newcommand{\etaO}{\eta_{\rm{O}}}
\newcommand{\etaA}{\eta_{\rm{A}}}

\newcommand{\gmag}{\gamma_{\rm{mag}}}
\newcommand{\gturb}{\gamma_{\rm{turb}}}

\newcommand{\msun}{\thinspace M_{\odot}}

\newcommand{\zsun}{\thinspace Z_{\odot}}
\newcommand{\gcm}{\thinspace \rm{g} \thinspace \rm{cm}^{-3}}
\newcommand{\gcmcms}{\thinspace \rm{g} \thinspace \rm{cm^2} \thinspace \rm{s}^{-1}}  

\newcommand{\magB}{\bm{B}}
\newcommand{\magBhat}{\bm{\hat{B}}}
\newcommand{\magmu}{\mu}

\newcommand{\vel}{\bm{v}}

\newcommand{\pvec}{\bm{p}_{\rm{pol}}}

\newcommand{\cms}{\thinspace \rm{cm} \thinspace \rm{s}^{-1}}

\newcommand{\kms}{\thinspace \rm{km} \thinspace \rm{s}^{-1}}

\usepackage{xcolor}
\definecolor{steelblue}{rgb}{0.275, 0.510, 0.706}
\definecolor{seagreen}{rgb}{0.190, 0.525, 0.361}
\definecolor{alcrimson}{rgb}{0.227, 0.38, 0.54}

\received{}
\revised{}
\accepted{}
\submitjournal{ApJ}

\shorttitle{Unipolar outflow and protostellar rocket effect}
\shortauthors{Takaishi et al.}

\begin{document}

\title{
  Formation of unipolar outflow and \textit{protostellar rocket effect}\\
  in magnetized turbulent molecular cloud cores
}

\correspondingauthor{Daisuke Takaishi}
\email{k3790238@kadai.jp}

\author[0000-0001-6273-805X]{Daisuke Takaishi}
\affiliation{Graduate School of Science and Engineering, Kagoshima
  University, 1-21-35 Korimoto, Kagoshima, Kagoshima 890-0065, Japan}
\author[0000-0001-6738-676X]{Yusuke Tsukamoto}
\affiliation{Graduate School of Science and Engineering, Kagoshima
  University, 1-21-35 Korimoto, Kagoshima, Kagoshima 890-0065, Japan}
\author[0000-0002-2902-4239]{Miyu Kido}
\affiliation{Graduate School of Science and Engineering, Kagoshima
  University, 1-21-35 Korimoto, Kagoshima, Kagoshima 890-0065, Japan}
\author[0000-0003-0845-128X]{Shigehisa Takakuwa}
\affiliation{Graduate School of Science and Engineering, Kagoshima
  University, 1-21-35 Korimoto, Kagoshima, Kagoshima 890-0065, Japan}
\affiliation{Academia Sinica Institute of Astronomy and Astrophysics,
  11F of Astronomy-Mathematics Building, AS/NTU, No. 1, Sec. 4,
  Roosevelt Road, Taipei 10617, Taiwan}
\author[0000-0002-4093-6925]{Yoshiaki Misugi}
\affiliation{Center for Computational Astrophysics, National
  Astronomical Observatory of Japan, 2-21-1 Osawa, Mitaka, Tokyo
  181-8588, Japan}
\author[0000-0003-0548-1766]{Yuki Kudoh}
\affiliation{Astronomical Institute, Tohoku University, 6-3 Aramaki
  Aoba-ku, Sendai, Miyagi 980-8578, Japan}
\author[0000-0002-4858-7598]{Yasushi Suto}
\affiliation{Department of Physics, The University of Tokyo, Tokyo
  113-0033, Japan}
\affiliation{Research Center for the Early Universe, School of
  Science, The University of Tokyo, Tokyo 113-0033, Japan}
\affiliation{Laboratory of Physics, Kochi University of Technology,
  Tosa Yamada, Kochi 782-8502, Japan}

\begin{abstract}
  Observed protostellar outflows exhibit a variety of asymmetrical
  features, including remarkable unipolar outflows and bending
  outflows.  Revealing the formation and early evolution of such
  asymmetrical protostellar outflows, especially the unipolar
  outflows, is essential for a better understanding of the star and
  planet formation because they can dramatically change the mass
  accretion and angular momentum transport to the protostars and
  protoplanetary disks.  Here, we perform the three-dimensional
  non-ideal magnetohydrodynamics simulations to investigate the
  formation and early evolution of the asymmetrical protostellar
  outflows in magnetized turbulent isolated molecular cloud cores.  We
  find, for the first time to our knowledge, that the unipolar outflow
  forms even in the single low-mass protostellar system.  The results
  show that the unipolar outflow is driven in the weakly magnetized
  cloud cores with the dimensionless mass-to-flux ratios of $\magmu=8$
  and $16$.  Furthermore, we find the \textit{protostellar rocket
    effect} of the unipolar outflow, which is similar to the launch
  and propulsion of a rocket.  The unipolar outflow ejects the
  protostellar system from the central dense region to the outer
  region of the parent cloud core, and the ram pressure caused by its
  ejection suppresses the driving of additional new outflows.  In
  contrast, the bending bipolar outflow is driven in the moderately
  magnetized cloud core with $\magmu=4$.  The ratio of the magnetic to
  turbulent energies of a parent cloud core may play a key role in the
  formation of asymmetrical protostellar outflows.
\end{abstract}

\keywords{Magnetic fields (994) --- Protoplanetary disks (1300) ---
  Protostars (1302) --- Star formation (1569) --- Stellar outflows
  (1636) --- Young stellar objects (1834)}

\section{Introduction}
\label{sec:intro}

Protostellar outflows play essential roles in the formation process of
protostars and planets.  The protostellar outflows are one of the
observable signs of the birth of protostars in the dense regions of
molecular cloud cores.  The mass accretion and angular momentum
transport to protostars and protoplanetary disks are regulated by the
driving and subsequent evolution of the protostellar outflows,
contributing to the determination of the star formation efficiency in
the isolated protostellar cores \citep{2013MNRAS.431.1719M}.
Moreover, the grown dust with a size of about a centimeter in the
inner region of the protoplanetary disk is entrained by the
protostellar outflow and refluxed from the outflow onto the outer
region of several tens of astronomical units of the disk
\citep{2021ApJ...920L..35T}.  The above process, which is referred to
as the \textit{ashfall} phenomenon, can circumvent the radial drift
barrier that typically hinders planet formation.  Therefore, the
protostellar outflows contribute to crucial physical processes for the
formation of protostars and planets.

Protostellar outflows are ubiquitously observed in star-forming
regions.  The outflows are typically traced by the high-velocity wings
of molecular line emissions \citep[e.g.,][]{1990ApJ...356..184M}.
Since the unexpected first discovery of a bipolar molecular outflow
with high-velocity $\rm{^{12}CO}$ wings driven by low-mass protostars
in L1551 IRS 5 \citep{1980ApJ...239L..17S}, a significant number of
bipolar molecular outflows driven by Young Stellar Objects (YSOs),
ranging from low-mass, intermediate-mass and high-mass YSOs as well as
young brown dwarfs and proto-brown-dwarf candidates, have been
observed \citep[e.g.,][]{1983ApJ...265..824B, 1985ARA&A..23..267L,
  1989Natur.342..161F, 1992A&A...261..274C, 1993prpl.conf..603F,
  1996A&A...311..858B, 2002A&A...383..892B, 2004A&A...426..503W,
  2006ApJ...646.1070A, 2007prpl.conf..245A, 2008ApJ...689L.141P,
  2013ApJ...774...22P, 2014ApJ...783...29D, 2019ApJS..240...18K,
  2020ApJ...896...11F, 2023ApJ...945...63H}.  In addition,
extragalactic molecular outflows driven by high-mass YSOs are
discovered in nearby low-metallicity galaxies, the Large Magellanic
Cloud with the metallicity of $Z=0.5\zsun$ \citep{2015ApJ...807L...4F,
  2016ApJ...827...72S, 2019ApJ...886...14F, 2019ApJ...886...15T,
  2022ApJ...933...20T} and the Small Magellanic Cloud with
$Z=0.2\zsun$ \citep{2022ApJ...936L...6T}.  The outflows are therefore
fundamental and universal feedback phenomena in the star formation
process.

One of the remarkable discoveries is that the molecular outflows
driven by YSOs exhibit a variety of asymmetrical features.  For
instance, a statistical study of the properties of molecular outflows
conducted by \citet{2004A&A...426..503W} show that 50 sources (13 \%)
out of 391 targets present unipolar outflows, of which 28 are
redshifted ones, indicating that the redshifted unipolar outflows are
equally abundant compared to the blueshifted ones and the unipolar
outflows are intrinsic.  Recent high-resolution observations of the
protostellar environment in the Orion A molecular cloud using the
Atacama Large Millimeter/submillimeter Array (ALMA) have more clearly
detected the isolated protostars driving the unipolar outflow
\citep{2023ApJ...947...25H}.  The resolved unipolar cavities in the
circumstellar envelopes are delineated by the near-infrared images of
the scattered light around protostars, which typically trace the hot
shocked gas regions in the outflows, obtained by the survey of
protostellar outflow cavities in the Orion molecular clouds using the
Hubble Space Telescope \citep{2021ApJ...911..153H}.  Although the
optical and near-infrared observations are inherently affected by the
extinction effect, the result can provide complementary evidence for
the existence of the unipolar outflows.  Additionally, many detections
of unipolar outflows in low-mass and high-mass YSOs are reported from
recent ALMA observations \citep[e.g.,][]{2018ApJ...863...19A,
  2018A&A...618A.120L, 2019ApJ...874..104K, 2019ApJ...887..209A,
  2020A&A...634L..12D, 2020A&A...636A..65G, 2020ApJ...903..119L,
  2020ApJS..251...20D, 2021MNRAS.507.4316B, 2023ApJ...944...92S,
  2023arXiv230615346D, 2023ApJ...958..102S}.  Moreover, observations
have frequently detected bipolar molecular outflows with asymmetrical
features, such as the bendings and different sizes of redshifted and
blueshifted lobes \citep{2013ApJ...774...39A, 2015ApJ...799..193Y,
  2017ApJ...834..178Y, 2018ApJ...863...19A, 2019ApJ...887..209A,
  2021ApJ...910...11O, 2023ApJ...947...25H, 2023ApJ...953..190K}.
These observational results suggest that the asymmetrical features
frequently emerge in the outflows.

Although the unipolar outflows have been observed, the formation and
early evolution of the unipolar outflows driven by low-mass protostars
remain unclear so far.  Understanding the formation and early
evolution of the outflows with the outstanding asymmetrical features,
in particular the unipolar outflows, is crucial because they can
strongly regulate the mass accretion and angular momentum transport to
the protostars and protoplanetary disks.  Observed molecular cloud
cores are often threaded by the magnetic field
\citep{2012ARA&A..50...29C, 2023ASPC..534..193P}, and in the context
of low-mass star formation, the bipolar outflows magnetically driven
by the first cores and the protoplanetary disks are successfully
reproduced by many numerical simulations of the gravitational collapse
of rotating magnetized cloud cores
\citep[e.g.,][]{1998ApJ...502L.163T, 2002ApJ...575..306T,
  2004ApJ...616..266M, 2006ApJ...641..949B, 2008ApJ...676.1088M,
  2010ApJ...714L..58T, 2013ApJ...763....6T, 2014MNRAS.437...77B,
  2015ApJ...801..117T, 2015MNRAS.452..278T, 2017PASJ...69...95T,
  2018MNRAS.475.1859W, 2018A&A...615A...5V, 2018ApJ...868...22T,
  2020ApJ...896..158T, 2020ApJ...898..118H, 2021ApJ...920L..35T,
  2021MNRAS.507.2354W, 2023A&A...670A..61M}.  On the other hand, the
isolated low-mass cloud cores are often associated with the turbulent
velocity field as well \citep[e.g.,][]{1993ApJ...406..528G,
  1998ApJ...504..207B, 2000ApJ...543..822B, 2019ApJ...881...11M,
  2023ApJ...943...76M}, which is typically subsonic or at most
transonic \citep{2007prpl.conf...33W}.  The turbulence in the low-mass
cloud cores can naturally produce the complex asymmetrical structures
such as the warped protoplanetary disks and the filamentary infalling
envelopes, and the rotation directions of the disks change dynamically
as time proceeds due to the chaotic accretions
\citep[e.g.,][]{2013MNRAS.428.1321T, 2020MNRAS.492.5641T,
  2021PASJ...73L..25T}.  Thus, the turbulence is expected to generate
the asymmetry of the magnetic field, leading to the unipolar outflows.
However, few studies have performed that the asymmetrical bipolar
outflows form in the realistic simulations of the collapse of
magnetized turbulent low-mass cloud cores with/without coherent
rotation \citep{2011ApJ...728...47M, 2013A&A...554A..17J,
  2017ApJ...839...69M, 2018MNRAS.477.4241L}, and they have not yet
identified the unipolar outflows driven by low-mass protostars.

In contrast to the above simulations, in a recent paper,
\citet{2021A&A...656A..85M} report that the weak and transient
unipolar outflow is driven by evolving binary massive protostars
formed in the most turbulent case of their simulations, which
calculate the gravitational collapse of magnetized turbulent massive
cores of $100\msun$ by solving the magnetohydrodynamics (MHD)
equations including the ambipolar diffusion and hybrid radiative
transfer in the context of high-mass star formation (see also
\citet{2021A&A...652A..69M} for details of their simulations).  The
results indicate that the initial turbulence of the parent cloud core,
i.e. the environmental ram pressure, can strongly affect the outflow
driving.  \citet{2020MNRAS.499.4490M} also show that the ram pressure
caused by the infalling envelope with the high accretion rate
suppresses the outflow driving when the rigidly rotating cloud core
initially has a weaker magnetic field.  These results suggest that the
ram pressure caused by the stronger turbulence and weaker magnetic
field may be crucial for the formation of the unipolar outflows in the
low-mass protostar formation as well as in the high-mass protostar
formation.

This paper reports, for the first time to our knowledge, the formation
of the unipolar outflows driven by the single low-mass protostellar
systems.  We perform the simulations of the gravitational collapse of
magnetized turbulent low-mass cloud cores of $1\msun$ with strong,
moderate and weak magnetic fields to investigate the formation and
early evolution of the unipolar outflows, which have not yet been
explored in the previous studies.  In addition, this paper presents
the subsequent evolution of the protostellar system driving the
unipolar outflow, which is similar to the launch and propulsion of a
rocket.

The rest of the paper consists of the following sections.  Section
\ref{sec:Method_and_IC} describes the numerical method and the initial
conditions.  Section \ref{sec:results} presents the numerical results.
Finally, Section \ref{sec:summary_and_discussion} summarizes and
discusses the results and findings.


\section{Method and Initial Conditions}
\label{sec:Method_and_IC}

\subsection{Basic Equations and Numerical Method}
\label{subsec:eqs-method}

The simulations solve the non-ideal MHD equations including
self-gravity of the gas:
\begin{align}
  \frac{D\vel}{Dt} &= -\frac{1}{\rho}
  \left\{
  \nabla\left(P+\frac{1}{2}|\magB|^2\right)
  -\nabla\cdot\left(\magB\magB\right)
  \right\}
  -\nabla\phi, \label{eq:motion} \\
  \frac{D\magB}{Dt} &=
  \left(\magB\cdot\nabla\right)\vel
  -\magB\left(\nabla\cdot\vel\right)
  -\nabla \times \left\{
  \etaO (\nabla \times \magB)
  -\etaA ( (\nabla\times\magB)\times\magBhat )
  \times \magBhat \right\}, \label{eq:induction} \\
  \nabla^2 \phi &= 4 \pi G \rho, \label{eq:poisson}
\end{align}
where $\vel$ is the gas velocity, $\rho$ is the gas density, $P$ is
the gas pressure, $\magB$ is the magnetic field, $\phi$ is the
gravitational potential, and $G$ is the gravitational constant.  The
unit vector along the magnetic field is denoted by
$\magBhat\equiv\magB/|\magB|$.  $\etaO$ and $\etaA$ are the
resistivities for the Ohmic dissipation and ambipolar diffusion of the
non-ideal MHD effects.  We note that the Hall effect, which is another
important factor of the non-ideal MHD effects, is currently ignored
because the simulation incorporating the Hall effect requires highly
small time steps in its calculation and thus it is computationally
highly demanding for the long-term simulation up to
$\sim10^{4}\thinspace\rm{yr}$ after the protostar formation.

Instead of solving the radiation transfer, we adopt the barotropic
equation of state that mimics the thermal evolution of the cloud core
presented by the radiation hydrodynamics simulations
\citep{2000ApJ...531..350M}:
\begin{equation}
  P(\rho) = c_{\rm{s,iso}}^2 \rho
  \left\{ 1+\left(\frac{\rho}{\rho_{\rm{crit}}}\right)^{2/3} \right\},
  \label{eq:eos}
\end{equation}
where $c_{\rm{s,iso}}=1.9\times10^4\cms$ is the isothermal sound speed
at the temperature of $10\thinspace\rm{K}$, and
$\rho_{\rm{crit}}=4\times10^{-14}\gcm$ is the critical density at
which the thermal evolution changes from the isothermal to adiabatic.

The equations are solved with the smoothed particle hydrodynamics
(SPH) method \citep{1977AJ.....82.1013L, 1977MNRAS.181..375G,
  1985A&A...149..135M}.  The ideal MHD part of the equations is solved
with the Godunov smoothed particle magnetohydrodynamics (GSPMHD)
method \citep{2011MNRAS.418.1668I}.  In addition, we adopt the
hyperbolic divergence cleaning method proposed by
\citet{2013ASPC..474..239I} so that the divergence-free condition of
the magnetic field is satisfied.  The Ohmic dissipation and ambipolar
diffusion are calculated with the methods of
\citet{2013MNRAS.434.2593T} and \citet{2014MNRAS.444.1104W}.  The
processes of the Ohmic dissipation and ambipolar diffusion are
accelerated by super-time-stepping (STS) method \citep{Alexiades1996}.
The parameters of $\nu_{\rm{sts}}=0.01$ and $N_{\rm{sts}}=5$ are
adopted for STS \citep{2013MNRAS.434.2593T}.  The self-gravity of the
gas is computed by the Barnes-Hut octree algorithm with an opening
angle parameter of $\theta_{\rm{gravity}}=0.5$
\citep{1986Natur.324..446B}.  The spline interpolation for the
gravitational softening is adopted with the technique of the adaptive
softening length \citep{2007MNRAS.374.1347P}.  The numerical code is
parallelized with the Message Passing Interface (MPI).  The numerical
code has already been applied to a variety of problems
\citep[e.g.,][]{2011MNRAS.416..591T, 2013MNRAS.428.1321T,
  2013MNRAS.434.2593T, 2013MNRAS.436.1667T, 2015MNRAS.446.1175T,
  2015ApJ...810L..26T, 2015MNRAS.452..278T, 2017PASJ...69...95T,
  2018ApJ...868...22T, 2020MNRAS.492.5641T, 2020ApJ...896..158T,
  2021ApJ...913..148T, 2021ApJ...920L..35T, 2021PASJ...73L..25T,
  2023PASJ...75..835T}.


\subsection{Resistivity model}
\label{subsec:resistivity_model}

The simulations use the tabulated resistivities for $\etaO$ and
$\etaA$ that are presented as the single-sized dust model with the
dust size of $a_{\rm{d}}=0.035\mum$ in \citet{2020ApJ...896..158T}.
The resistivities are generated by the chemical reaction network
calculation using the method of \citet{2015ApJ...801...13S}.  The
calculation includes the ion species of $\rm{H^+}$, $\rm{H_2^+}$,
$\rm{H_3^+}$, $\rm{HCO^+}$, $\rm{Mg^+}$, $\rm{He^+}$, $\rm{C^+}$,
$\rm{O^+}$, $\rm{O_2^+}$, $\rm{H_3O^+}$, $\rm{OH^+}$, and
$\rm{H_2O^+}$, and the neutral species of $\rm{H}$, $\rm{H_2}$,
$\rm{He}$, $\rm{CO}$, $\rm{O_2}$, $\rm{Mg}$, $\rm{O}$, $\rm{C}$,
$\rm{HCO}$, $\rm{H_2O}$, $\rm{OH}$, $\rm{N}$, and $\rm{Fe}$.  The
calculation also includes the neutral and singly charged dust grains,
$\rm{G^{0}}$, $\rm{G^{+}}$, and $\rm{G^{-}}$.  The calculation takes
into account the cosmic-ray ionization, gas-phase and dust-surface
recombination, and ion-neutral reactions.  The indirect ionization by
high-energy photons emitted by direct cosmic-ray ionization (described
as CRPHOT in the UMIST database) is also considered in the
calculation.  The initial abundance and reaction rates are taken from
the UMIST2012 database \citep{2013A&A...550A..36M}.  The grain-ion and
grain-grain collision rates are calculated using the equations in
\citet{1987ApJ...320..803D}.  The chemical reaction network
calculation is conducted using the CVODE package \citep{Hindmarsh2005}
assuming the system is in the chemical equilibrium.  The resistivities
are calculated using the abundances of charged species in the
equilibrium state.  The momentum transfer rate between the charged and
neutral species is calculated using the equations in
\citet{2008A&A...484...17P}.  The temperature for the chemical
reaction network calculation is modeled as
$T_{\rm{chem}}=10(1+\gamma_{\rm{T}}(\rho/\rho_{\rm{crit}})^{(\gamma_{\rm{T}}-1)})
  \thinspace\rm{K}$, where $\gamma_{\rm{T}}=7/5$.  It is assumed that
  the dust has the internal density of $\rho_{\rm{d}}=2\gcm$ and the
  size of $a_{\rm{d}}=0.035\mum$.  The dust-to-gas mass ratio is fixed
  to be 0.01 in the calculation.  The cosmic ray ionization rate is
  assumed to be $\xi_{\rm{CR}}=10^{-17}\thinspace\rm{s^{-1}}$.


\subsection{Initial conditions for the density profile}
\label{subsec:init_cond}

The simulations start from the collapse of isolated molecular cloud
cores including the magnetic field and turbulence simultaneously.  The
initial cloud core follows the density profile presented by
\citet{2020ApJ...896..158T}:
\begin{eqnarray}
  \rho(r) &=& \left\{
  \begin{array}{ll}
    f\rho_{\rm{c}}\varrho_{\rm{BE}}\left(r/a\right) & \rm{for}~{\it{r<R_{\rm{c}}}} \\
    f\rho_{\rm{c}}\varrho_{\rm{BE}}\left(R_{\rm{c}}/a\right)
    \left(\frac{r}{R_{\rm{c}}}\right)^{-4} & \rm{for}~{\it{R_{\rm{c}} \leq r<{\rm 10}R_{\rm{c}}}}
  \end{array}
  \right., \label{eq:prof_rho} \\
  a &\equiv& c_{\rm{s,iso}} \left( \frac{1}{4\pi G \rho_{\rm{c}}} \right)^{1/2}, \label{eq:prof_a}
\end{eqnarray}
where $f$ is the density enhancement factor, $\rho_{\rm{c}}$ is the
characteristic density, and $R_{\rm{c}}=6.45a$ is the radius of the
initial cloud core.  $\varrho_{\rm{BE}}$ is the non-dimensional
density profile of the Bonnor-Ebert sphere \citep{1956MNRAS.116..351B,
  1955ZA.....37..217E}, which is a pressure-confined and
self-gravitating isothermal gas sphere in hydrostatic equilibrium
state against the gravitational collapse and well fits the density
distribution of observed isolated molecular cloud cores
\citep[e.g.,][]{2001Natur.409..159A, 2005AJ....130.2166K}.  The
density enhancement factor $f$ controls the strength of gravity.  More
specifically, $f$ can be written as $f=0.84/\alpha$, where
\begin{equation}
  \alpha = \frac{E_{\rm{thm}}}{|E_{\rm{grav}}|}, \label{eq:alpha}
\end{equation}
with $E_{\rm{thm}}$ and $E_{\rm{grav}}$ are the thermal and
gravitational energies of the initial cloud core (without the
surrounding medium of $\rho(r)\propto r^{-4}$); see Appendix A of
\citet{2011ApJ...728...47M}.  The density profile of the initial cloud
core with $f=1$ in a region of $r<R_{\rm{c}}$ corresponds to that of
the critical Bonnor-Ebert sphere.  The initial cloud core with $f>1$
is gravitationally unstable.

The initial cloud core has the temperature of
$T_{\rm{c}}=10\thinspace\rm{K}$, the radius of
$R_{\rm{c}}=4.8\times10^3\thinspace\rm{au}$, the mass of
$M_{\rm{c}}=1\msun$ within $r<R_{\rm{c}}$ ($\sim2.1\msun$ in the
entire domain of $r<10R_{\rm{c}}$), and the ratio of $\alpha=0.4$,
which are determined by specifying $f=2.1$ and the central density of
$\rho_{0}=f\rho_{\rm{c}}=7.3\times10^{-18}\gcm$.  The free-fall time
is calculated as
$t_{\rm{ff}}=(3\pi/(32G\rho_{0}))^{1/2}=2.5\times10^{4}\thinspace\rm{yr}$.
The simulations resolve $M_{\rm{c}}=1\msun$ with the number of SPH
particles of $N_{\rm{SPH}}=10^6$, which corresponds to the mass
resolution of $M_{\rm{c}}/N_{\rm{SPH}}=10^{-6}\msun$.  The number of
all the SPH particles in the entire domain of $r<10R_{\rm{c}}$ is
$\sim 2.1\msun/10^{-6}\msun=2.1\times10^{6}$.


\subsection{Turbulence and magnetic fields}
\label{subsec:turb_and_mag}

The initial cloud core has the divergence-free turbulent velocity
field with the velocity power spectrum of $P_{v}(k) \propto k^{-4}$
\citep{2000ApJ...543..822B}, where $k$ is the wavenumber.  The
amplitude of the turbulence is characterized by the mean sonic Mach
number:
\begin{equation}
  {\cal{M}}_{\rm{s}} = \frac{\sqrt{3}\sigma_{\rm{v}}}{c_{\rm{s,iso}}}, \label{eq:Ms}
\end{equation}
or the ratio of the turbulent to gravitational energies:
\begin{equation}
  \gturb = \frac{E_{\rm{turb}}}{|E_{\rm{grav}}|}, \label{eq:gturb}
\end{equation}
where $\sigma_{\rm{v}}$ and
$E_{\rm{turb}}=3\sigma_{\rm{v}}^2M_{\rm{c}}/2$ are the one-dimensional
velocity dispersion and turbulent energy of the initial cloud core
without the surrounding medium of $\rho(r)\propto r^{-4}$.  We adopt
${\cal{M}}_{\rm{s}}=0.86$, which corresponds to $\gturb=0.1$
initially.  Hence, the initial cloud core has a non-vanishing net
angular momentum of $|\bm{J}_{\rm{c,net}}|=4.4\times10^{53}\gcmcms$
due to the stochastic nature of the turbulent velocity field.  The
direction of $\bm{J}_{\rm{c,net}}$ is set as the $z$-axis of the
simulations.

The turbulent velocity field is generated with the method of
\citet{2013MNRAS.428.1321T}, which is also adopted in our previous
studies \citep{2020MNRAS.492.5641T, 2021PASJ...73L..25T}.  All the
simulations use the same turbulent velocity field, which is assigned
to the cloud core of $r<R_{\rm{c}}$ alone, and vanishes for
$R_{\rm{c}} \leq r <10R_{\rm{c}}$.  We note that the initial velocity
field consists only of the turbulent velocity field, and it is not the
superposition of the turbulent and rotational velocity fields.

The initial cloud core has the axisymmetric magnetic field modeled by
\citet{2020ApJ...896..158T}.  In cylindrical coordinates
$(R,\thinspace \varphi,\thinspace z)$, they are
\begin{eqnarray}
  B_{R}(R,z) &=& \frac{B_{\rm{c}}}
  {\left[1+\left(R/R_{\rm{c}}\right)^2+\left(z/R_{\rm{c}}\right)^2\right]^2}
  \left(\frac{R}{R_{\rm{c}}}\right)
  \left(\frac{z}{R_{\rm{c}}}\right), \label{eq:BR} \\
  B_{z}(R,z) &=& \frac{B_{\rm{c}}}
  {\left[1+\left(R/R_{\rm{c}}\right)^2+\left(z/R_{\rm{c}}\right)^2\right]^2}
  \left\{1+\left(\frac{z}{R_{\rm{c}}}\right)^2\right\}, \label{eq:Bz}
\end{eqnarray}
where $B_{\rm{c}}$ denotes the strength of the central magnetic field.
The magnetic field described by the equations has a constant uniform
component for $z$-direction in the central region ( $B_{R}\rightarrow
0$ and $B_{z}\rightarrow B_{\rm{c}}$ as $r\rightarrow 0$) and becomes
an hourglass-shaped structure with $|\magB| \propto r^{-2}$ in the
outer region (except at the midplane) as $r \rightarrow \infty$ (see
Appendix A of \citet{2020ApJ...896..158T} for details).  The
simulations can avoid to emerge low-$\beta_{\rm{plasma}}$ regions in
the surrounding medium of $\rho(r)\propto r^{-4}$, where
$\beta_{\rm{plasma}}$ is the plasma beta parameter.  Such the
hourglass-shaped structures of the magnetic field have been estimated
by previous observations on protostellar cores
\citep[e.g.,][]{2006Sci...313..812G, 2020ApJ...891...55K} and starless
cores \citep[e.g.,][]{2017ApJ...845...32K, 2018ApJ...865..121K,
  2020PASJ...72....8K}.

The initial cloud cores are parameterized with the strength of the
central magnetic field $B_{\rm{c}}$, which can be expressed by using
the dimensionless mass-to-flux ratio:
\begin{eqnarray}
  \magmu &=&
  \left(\frac{M_{\rm{c}}}{\Phi_{\rm{mag}}\left(R_{\rm{c}}\right)}\right) \left/
  \left(\frac{M_{\rm{c}}}{\Phi_{\rm{mag}}}\right)_{\rm{crit}} \right. , \label{eq:mass-to-flux}
\end{eqnarray}
where $\Phi_{\rm{mag}}\left(R\right)$ is the magnetic flux of the
initial cloud core calculated as
\begin{eqnarray}
  \Phi_{\rm{mag}}\left( R \right) &=&
  \frac{\pi R^2 B_{\rm{c}}}{\left(R/R_{\rm{c}}\right)^2+1}, \label{eq:mag_flux}
\end{eqnarray}
and $(M_{\rm{c}}/\Phi_{\rm{mag}})_{\rm{crit}}=(0.53/3\pi)(5/G)^{1/2}$
is the critical mass-to-flux ratio on stability for uniform spheres
\citep{1976ApJ...210..326M}.

The simulations are conducted with four different dimensionless
mass-to-flux ratios, $\magmu=2,\thinspace 4,\thinspace 8,$ and $16$,
which correspond to $B_{\rm{c}}=252\thinspace\mu\rm{G}$,
$126\thinspace\mu\rm{G}$, $63\thinspace\mu\rm{G}$, and
$31\thinspace\mu\rm{G}$, respectively.  The dimensionless mass-to-flux
ratio with the constant value of the central magnetic field
$B_{\rm{c}}$ is defined as
\begin{eqnarray}
  \magmu_{\rm{const}} =
  \left( \frac{M_{\rm{c}}}{\pi R_{\rm{c}}^2B_{\rm{c}}} \right) \left/
  \left(\frac{M_{\rm{c}}}{\Phi_{\rm{mag}}}\right)_{\rm{crit}} \right.
  = \frac{\magmu}{2}. \label{eq:mu_const}
\end{eqnarray}
As noted in \citet{2020ApJ...896..158T}, $\magmu_{\rm{const}}$ would
be a suitable indicator to compare the strength of the magnetic field
of this study with those of previous studies because we focus on the
time evolution of the central region of the cloud core.  $\magmu$ can
be converted into the ratio of the magnetic to gravitational energies:
\begin{eqnarray}
  \gmag = \frac{E_{\rm{mag}}}{|E_{\rm{grav}}|}, \label{eq:gmag}
\end{eqnarray}
where $E_{\rm{mag}}$ is the magnetic energy of the initial cloud core
without the surrounding medium of $\rho(r)\propto r^{-4}$.

The model names and corresponding parameters are summarized in Table
\ref{table:summary_parameters}.  We note that the last column of Table
\ref{table:summary_parameters} summarizes the results of driven
outflow morphologies.

\begin{deluxetable*}{ccccccc}
  \tablecaption{
    The model names and parameters \label{table:summary_parameters}
  }
  \tablehead{
    \colhead{Model} &  
    \colhead{$\magmu$} &  
    \colhead{$B_{\rm{c}}\thinspace(\mu\rm{G})$} &  
    \colhead{$\magmu_{\rm{const}}$} &  
    \colhead{$\gmag$} &  
    \colhead{$E_{\rm{mag}}/E_{\rm{turb}}$} &  
    \colhead{Outflow}  
  }
  \startdata
  MF2  & 2  & 252 & 1 & 0.42   & 4.2   & No        \\
  MF4  & 4  & 126 & 2 & 0.10   & 1.0   & Bipolar   \\
  MF8  & 8  & 63  & 4 & 0.026  & 0.26  & Unipolar  \\
  MF16 & 16 & 31  & 8 & 0.0065 & 0.065 & Unipolar  \\
  \enddata
  \tablecomments{
    $\magmu$ is the dimensionless mass-to-flux ratio.
    $B_{\rm{c}}$ is the strength of the central magnetic field.
    $\magmu_{\rm{const}}$ is the dimensionless mass-to-flux ratio with
    the constant value of the central magnetic field.
    $\gmag=E_{\rm{mag}}/|E_{\rm{grav}}|$ is the ratio of the magnetic to gravitational energies.
    $E_{\rm{mag}}/E_{\rm{turb}}=\gmag/\gturb$ is the ratio of the magnetic to turbulent energies.
    The last column indicates the morphology of the driven outflow.
  }
\end{deluxetable*}


\section{Results}
\label{sec:results}

\subsection{overview on 3D structure of the driven outflows and magnetic field}
\label{subsec:overview}

\begin{figure*}
  \begin{center}
    \includegraphics[clip,width=180mm]{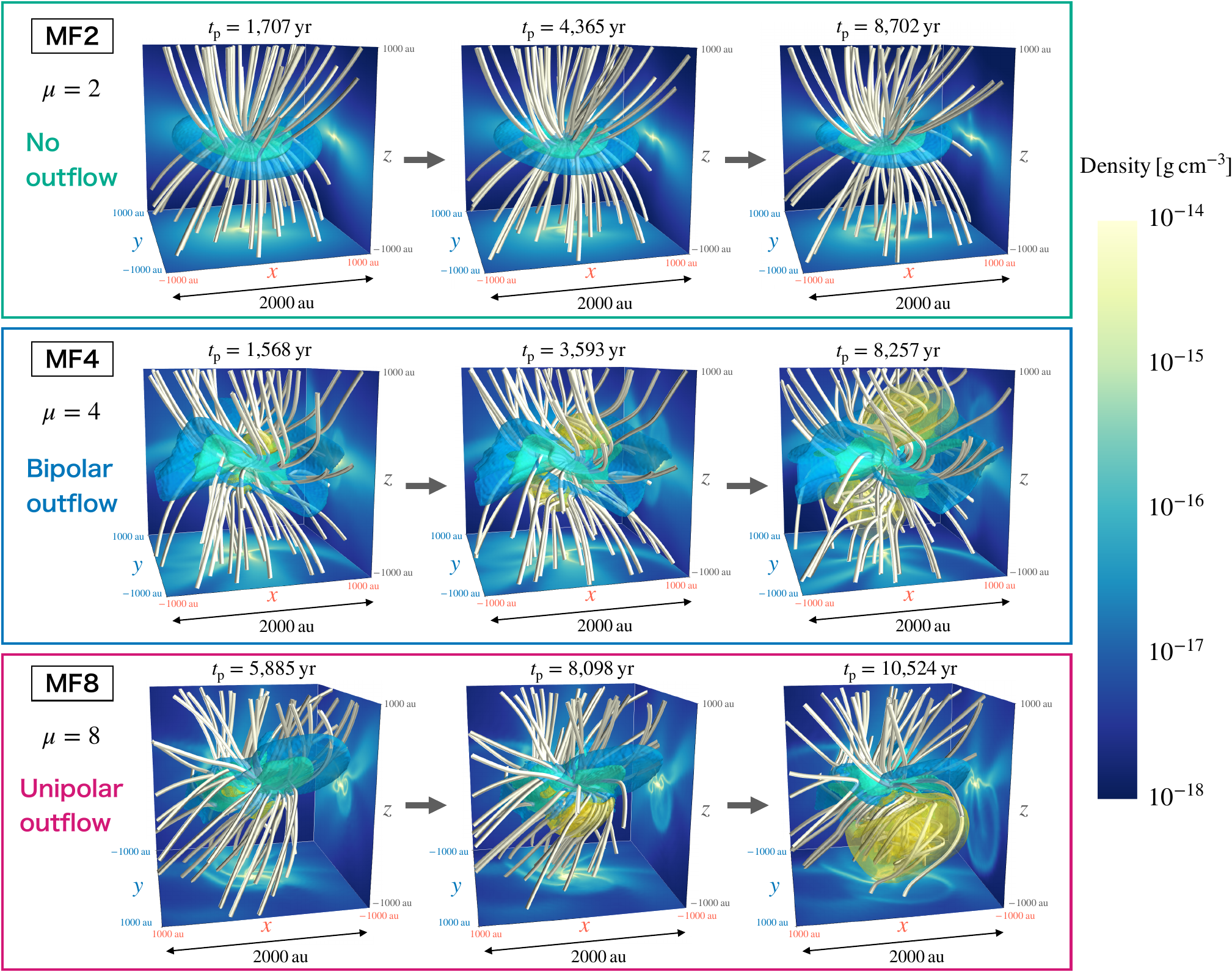}
  \end{center}
  \caption{Three-dimensional views of the density distribution and
    magnetic field structure for models of the no outflow (model MF2,
    top green box), bipolar outflow (model MF4, middle blue box), and
    unipolar outflow (model MF8, bottom magenta box).  Yellow
    isosurfaces show the density of $3.2\times10^{-18}\gcm$,
    $10^{-17}\gcm$, $3.2\times10^{-17}\gcm$, and $10^{-16}\gcm$ with
    the radial velocity of $v_r>0$, representing the outflows.  White
    lines show the magnetic field lines.  Blue and green isosurfaces
    show the density of $3.2\times10^{-17}\gcm$ and $10^{-16}\gcm$
    with $v_r<0$, representing the infalling envelopes.  Cut-plane
    densities on $x-y$ ($z=0$), $x-z$ ($y=0$), and $y-z$ ($x=0$)
    planes are projected for each panel.  $t_{\rm{p}}$ notes the
    elapsed time after the protostar formation epoch defined at the
    time when the central density becomes higher than
    $1.0\times10^{-11}\gcm$.  The scale of the box is
    $\sim2,000\thinspace\rm{au}$.  The origin of a coordinate system
    is shifted to the center of mass of the system.
    \label{fig:3D_comp_2000au}
  }
\end{figure*}

First, we show an overview of the simulations, which clearly show the
very different morphologies of the driven outflows.  Figure
\ref{fig:3D_comp_2000au} shows the three-dimensional views of the
density distribution and magnetic field structure for the three driven
outflows: the no outflow (model MF2, top green box), bipolar outflow
(model MF4, middle blue box), and unipolar outflow (model MF8, bottom
magenta box).  The simulations are conducted from the protostellar
collapse to $t_{\rm{p}}\sim10^4\thinspace\rm{yr}$, where $t_{\rm{p}}$
denotes the elapsed time after the protostar formation epoch defined
at the time when the central density becomes higher than
$1.0\times10^{-11}\gcm$.

The top green box of Figure \ref{fig:3D_comp_2000au} shows that no
outflow is driven in the model MF2 in which the initial cloud core has
a strong magnetic field of $\magmu=2$.  The hourglass-shaped structure
of the magnetic field is formed and kept until the end of the
simulation.  The protoplanetary disk at the central region does not
form in a timescale of $t_{\rm{p}}\sim~1.5\times10^4\thinspace\rm{yr}$
whereas the disk-shaped flattened infalling envelope, which is
so-called the pseudo-disk structure \citep{1993ApJ...417..220G,
  1993ApJ...417..243G}, forms in this model.  This result indicates
that the relatively strong magnetic field rapidly extracts the angular
momentum caused by the turbulent accretion from the central region via
magnetic braking and suppresses the formation of a Keplerian rotating
disk at the central region.

The middle blue box of Figure \ref{fig:3D_comp_2000au} shows that the
bipolar outflow is driven in the model MF4 in which the initial cloud
core has the magnetic field of $\magmu=4$.  The bipolar outflow is
mainly driven by the magneto-centrifugal wind model
\citep{1982MNRAS.199..883B} whereas the spiral-flow model
\citep{2011ApJ...728...47M, 2017ApJ...839...69M} also contributes to
driving it.  The magnetic field gradually becomes twisted as time
proceeds, and the driven bipolar outflow evolves to be a large size of
$\sim5\times10^2\thinspace\rm{au}$ in the lower region and
$\sim10^3\thinspace\rm{au}$ in the upper region, indicating that the
outflow size of the upper region is greater than that of the lower
region.  Furthermore, the results show that the driven bipolar outflow
is slightly bending and its driving directions in the upper and lower
regions are not perfectly antiparallel with each other.  The driving
directions of the bipolar outflow are also misaligned with the
direction of the large-scale global magnetic field (roughly the
$z$-axis direction).  The warped structures of the infalling envelopes
are emerged by the perturbation of the turbulent accretion.  Our
results suggest that the bending and misaligned structures of the
bipolar outflow are created by the turbulent accretion via the
surrounded infalling envelopes.  Therefore, the turbulence of
molecular cloud cores can naturally explain the observed asymmetrical
features of bipolar molecular outflows such as the bendings and
different sizes of redshifted and blueshifted lobes
\citep{2013ApJ...774...39A, 2015ApJ...799..193Y, 2017ApJ...834..178Y,
  2018ApJ...863...19A, 2019ApJ...887..209A, 2021ApJ...910...11O,
  2023ApJ...947...25H, 2023ApJ...953..190K}.  The results perform that
the bipolar outflows can be formed in not only the rotating cloud
cores but also the turbulent cloud cores, which is consistent with
previous theoretical studies \citep{2017ApJ...839...69M}.

The bottom magenta box of Figure \ref{fig:3D_comp_2000au} shows that
the unipolar outflow is driven in the model MF8 in which the initial
cloud core has the weak magnetic field of $\magmu=8$.  The unipolar
outflow evolves to be a large size of $\sim10^3\thinspace\rm{au}$ in
the lower region and has the large opening angle in the bottom right
panel of Figure \ref{fig:3D_comp_2000au}.  The magnetic field lines
gradually becomes twisted only in the lower region as time proceeds.
In contrast, the magnetic field lines in the upper region have
extended and spread out structures without twisting.  The results
indicate that the unipolar outflow is mainly driven by getting
accelerated with the magnetic pressure gradient force although the
magneto-centrifugal force \citep{1982MNRAS.199..883B} slightly
contributes to driving and accelerating it.
\citet{2002ApJ...575..306T} indeed reports that the outflow driven by
the magnetic pressure gradient force appears in the case of the weak
magnetic field while the outflow driven by the magneto-centrifugal
force appears in the case of the strong magnetic field because the
toroidal components of the magnetic field can be amplified more easily
by the disk rotation comparing to the case of the strong magnetic
field.  The results indicate that the strong turbulent accretions can
locally amplify the toroidal components of the magnetic field relative
to the poloidal ones, in contrasts to the bipolar outflow formed in
model MF4.  The results also show that the infalling envelopes have
warped and elongated filamentary structures by the turbulent
accretion.  As shown in Figure
\ref{fig:surface_density_2000au_mag_mu8OA} of the next subsection
\ref{subsec:unipolar_outflow}, these structures generated by the
turbulent accretion can naturally explain the observed arc-like
structures on the scale of $\sim10^3\thinspace\rm{au}$
\citep[e.g.,][]{2014ApJ...789L...4T}.

The unipolar outflow is also driven in the model MF16 in which the
initial cloud core has the very weak magnetic field of $\magmu=16$.
The outflows and initial parameters of the cloud cores are summarized
in Table \ref{table:summary_parameters}.  It can be seen from Table
\ref{table:summary_parameters} that the unipolar outflows are driven
with $E_{\rm{mag}}/E_{\rm{turb}}<1$, whereas the bipolar outflow is
driven with $E_{\rm{mag}}/E_{\rm{turb}}\sim1$.  The results of Figure
\ref{fig:3D_comp_2000au} and Table \ref{table:summary_parameters}
suggest that the ratio of the magnetic and turbulent energies of the
parent cloud core, as represented by $E_{\rm{mag}}/E_{\rm{turb}}$, may
play a key role in the driven outflow morphologies.


\subsection{formation and evolution of unipolar outflow}
\label{subsec:unipolar_outflow}

\begin{figure*}
  \begin{center}
    \includegraphics[clip,width=155mm]{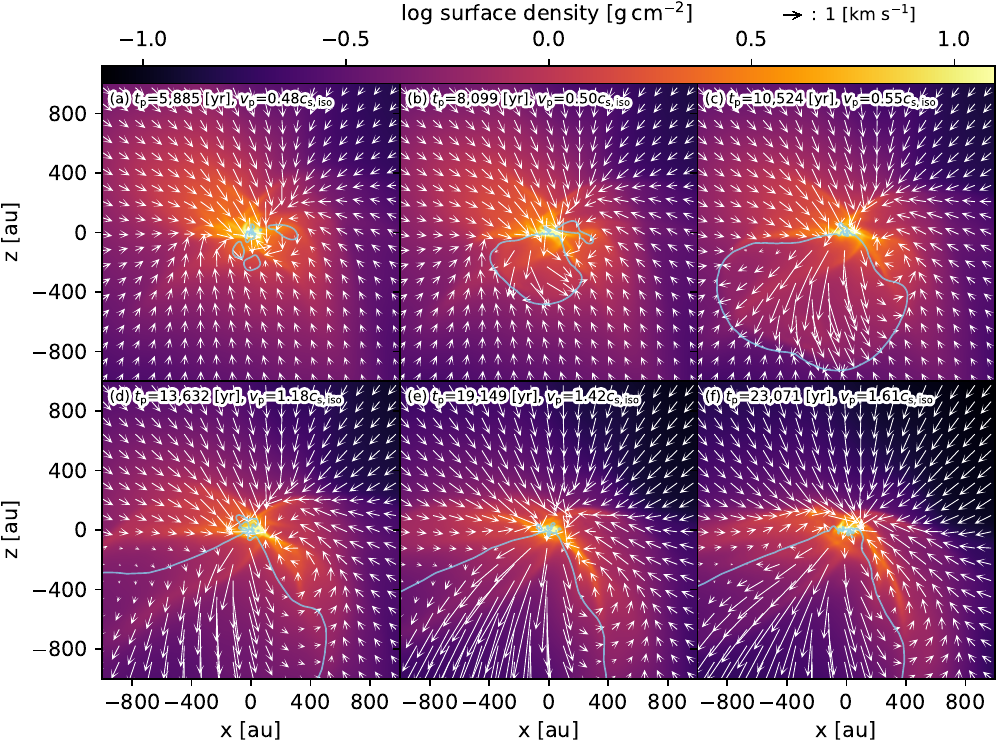}
  \end{center}
  \caption{Evolution of the surface density distributions along the
    $y$-direction for model MF8 in which the unipolar outflow is
    driven around the protostar.  Panels are labeled by $t_{\rm{p}}$
    that notes the elapsed time after the protostar formation epoch
    defined at the time when the central density becomes higher than
    $1.0\times10^{-11}\gcm$.  $v_{\rm{p}}$ is the velocity of the
    protostar.  White arrows of the panels show the cut-plane velocity
    at $y=0$.  The reference arrow plotted on the top right
    corresponds to $1\thinspace\kms$.  Sky-blue solid lines show the
    velocity contours at $v_r=0$, tracing the front lines of the
    outflowing gas, where $v_r$ is the radial velocity of the gas.
    The origin of a coordinate system is shifted to the center of mass
    of the system.
    \label{fig:surface_density_2000au_mag_mu8OA}
  }
\end{figure*}

Next, we focus on the formation and subsequent evolution of the
unipolar outflow.  Figure \ref{fig:surface_density_2000au_mag_mu8OA}
shows the evolution of the surface density distributions along the
$y$-direction for model MF8.  We note that the origin of a coordinate
system is shifted to the center of mass of the system.

Panel (b) of Figure \ref{fig:surface_density_2000au_mag_mu8OA} shows
that the unipolar outflow is driven around the protostar at
$t_{\rm{p}}\sim8\times10^3\thinspace\rm{yr}$ after the protostar
formation.  Subsequent evolution from panels (b) to (c), the unipolar
outflow grows up with the outflow speed of $1-5\kms$ and expands to
the scale of $\sim10^3\thinspace\rm{au}$ in a timescale of
$\sim5\times10^3\thinspace\rm{yr}$.  Panels (c) to (f) show that the
unipolar outflow has a highly wide opening angle.  The velocity of the
unipolar outflow gradually becomes large during the evolution from
panels (b) to (f).  As time proceeds from panels (b) to (f), the
velocity of the central protostar $v_{\rm{p}}$ evolves to be from
subsonic ($v_{\rm{p}}\sim 0.5c_{\rm{s,iso}}$) to supersonic
($v_{\rm{p}}\sim 1.6c_{\rm{s,iso}}$).

All the panel of Figure \ref{fig:surface_density_2000au_mag_mu8OA}
shows that arc-like structures appear from $\sim10^2\thinspace\rm{au}$
to $\sim10^3\thinspace\rm{au}$ scales in the $x-z$ plane.  The
arc-like structures in the panels of Figure
\ref{fig:surface_density_2000au_mag_mu8OA} are infalling by the
turbulent accretion and not outflowing.  \citet{2014ApJ...789L...4T}
have reported an arc-like structure on $\sim10^{3}\thinspace\rm{au}$
scale around a very low-luminosity protostar in the dense cloud core
MC27/L1521F by ALMA observations.  Many interferometric observations
have also detected similar arc-like structures, which have been
recently called the accretion streamers, in a wide range from the
cloud core scale of $\sim10^{4}\thinspace\rm{au}$
\citep[e.g.,][]{2020NatAs...4.1158P} to the disk and envelope scales
of $10^{2}-10^{3}\thinspace\rm{au}$
\citep[e.g.,][]{2014ApJ...793....1Y, 2017A&A...608A.134Y,
  2019AJ....157..165A, 2019ApJ...880...69Y, 2022ApJ...925...32T,
  2022A&A...658A.104G, 2023ApJ...953..190K, 2023ApJ...954..101A}.  Our
results suggest that the accretion streamers on the scales of
$10^{2}-10^{3}\thinspace\rm{au}$ can be naturally explained by the
filamentary envelope accretions caused by the turbulence of the parent
cloud cores, as emerged in many previous numerical simulations of the
collapse of self-gravitating low-mass cloud cores with the turbulence
\citep[e.g.,][]{2011ApJ...728...47M, 2013MNRAS.428.1321T,
  2017ApJ...839...69M, 2020MNRAS.492.5641T}.

\begin{figure*}
  \begin{center}
    \includegraphics[clip,width=155mm]{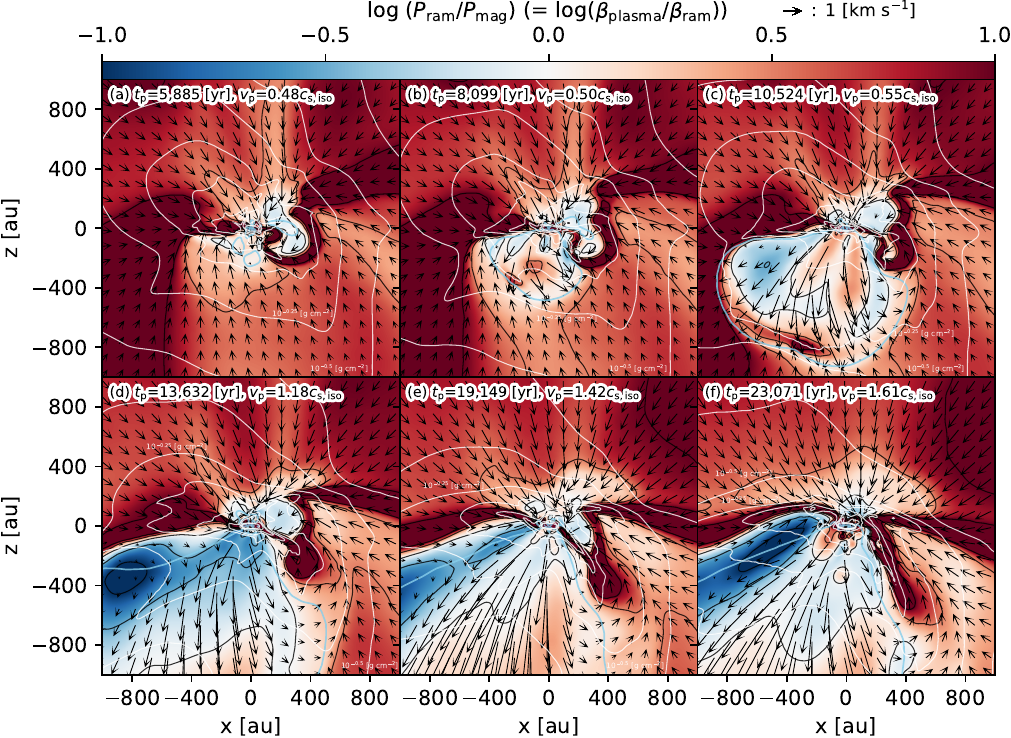}
  \end{center}
  \caption{Evolution of the ratio of the ram pressure $P_{\rm{ram}}$
    and magnetic pressure $P_{\rm{mag}}$ at the cut-plane of $y=0$ for
    model MF8.  Black lines show the contours of the ratio.
    $t_{\rm{p}}$ of each panel corresponds to that in Figure
    \ref{fig:surface_density_2000au_mag_mu8OA}.  $v_{\rm{p}}$ is the
    velocity of the protostar.  White lines show the contours of the
    surface density in Figure
    \ref{fig:surface_density_2000au_mag_mu8OA} ranging from $-0.75$ to
    $1.25$ in $0.25$ steps on a logarithmic scale.  Black arrows of
    the panels show the cut-plane velocity at $y=0$.  The reference
    arrow plotted on the top right corresponds to $1\thinspace\kms$.
    Sky-blue solid lines show the velocity contours at $v_r=0$,
    tracing the front lines of the outflowing gas.  The origin of a
    coordinate system is shifted to the center of mass of the system.
    \label{fig:ratio_PramPmag_2000au_mag_mu8OA}
  }
\end{figure*}

\begin{figure*}
  \begin{center}
    \includegraphics[clip,width=155mm]{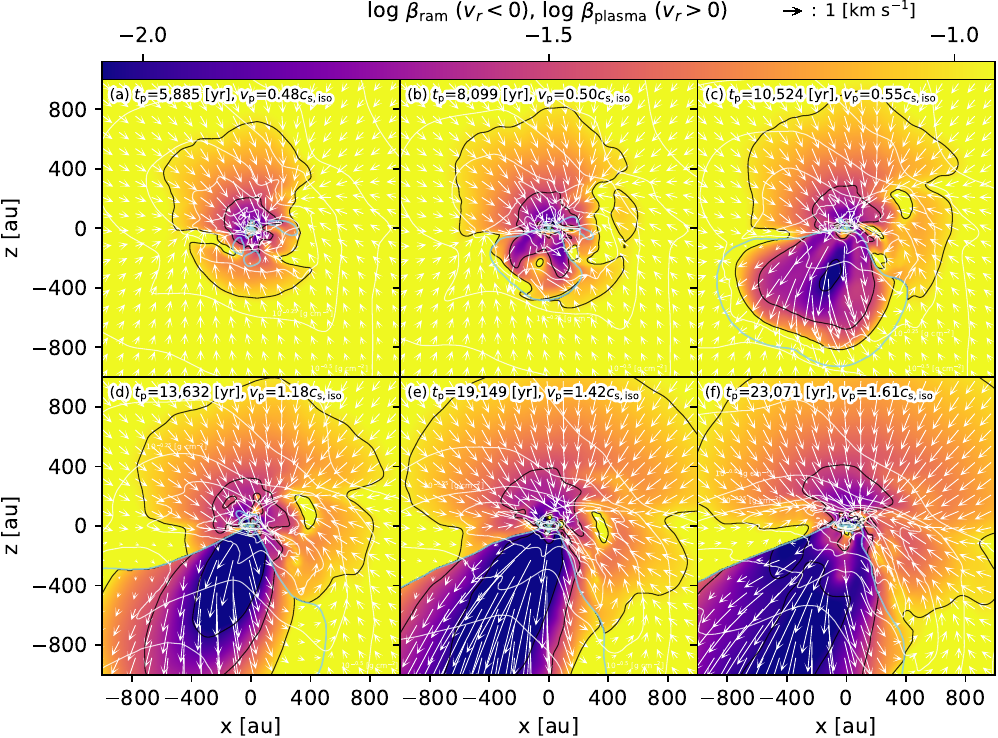}
  \end{center}
  \caption{Evolution of the ram beta parameter
    $\beta_{\rm{ram}}=P_{\rm{thm}}/P_{\rm{ram}}$ outside the outflow
    ($v_r<0$) and the plasma beta parameter
    $\beta_{\rm{plasma}}=P_{\rm{thm}}/P_{\rm{mag}}$ inside the outflow
    ($v_r>0$) at the cut-plane of $y=0$ for model MF8, where
    $P_{\rm{thm}}$ is the thermal pressure of the gas.  Black lines
    show the contours of them.  Sky-blue solid lines show the velocity
    contours at $v_r=0$, indicating the boundary between outside and
    inside of the outflow.  $t_{\rm{p}}$ of each panel corresponds to
    that in Figure \ref{fig:surface_density_2000au_mag_mu8OA}.
    $v_{\rm{p}}$ is the velocity of the protostar.  White lines show
    the contours of the surface density in Figure
    \ref{fig:surface_density_2000au_mag_mu8OA} ranging from $-0.75$ to
    $1.25$ in $0.25$ steps on a logarithmic scale.  White arrows of
    the panels show the cut-plane velocity at $y=0$.  The reference
    arrow plotted on the top right corresponds to $1\thinspace\kms$.
    The origin of a coordinate system is shifted to the center of mass
    of the system.
    \label{fig:beta_ram_beta_mag_2000au_mag_mu8OA}
  }
\end{figure*}

Figure \ref{fig:ratio_PramPmag_2000au_mag_mu8OA} shows the evolution
of the ratio of the ram pressure $P_{\rm{ram}}$ and magnetic pressure
$P_{\rm{mag}}$ at the cut-plane of $y=0$ for model MF8, following the
analysis of \citet{2020MNRAS.499.4490M}.  Panels (a) to (f) of Figure
\ref{fig:ratio_PramPmag_2000au_mag_mu8OA}, the ram pressure
$P_{\rm{ram}}$ is lower than the magnetic pressure $P_{\rm{mag}}$
inside the lower region of the driven unipolar outflow (inside the
sky-blue line).  The result indicates that the magnetic pressure
gradient force caused by the twisted magnetic field as shown in panels
of model MF8 in Figure \ref{fig:3D_comp_2000au} gradually becomes
large and continuously enhances the unipolar outflow getting
accelerated until the end of the simulation of
$t_{\rm{p}}\sim2.3\times10^4\thinspace\rm{yr}$.  In the upper region,
the ram pressure $P_{\rm{ram}}$, however, is always higher than the
magnetic pressure $P_{\rm{mag}}$ from panels (a) to (f) of Figure
\ref{fig:ratio_PramPmag_2000au_mag_mu8OA}, suggesting that the outflow
driving is suppressed and failed by the ram pressure due to the
infalling envelopes.

In order to see the detailed evolution of the ram and magnetic
pressures separately, we introduce the ram beta parameter as
$\beta_{\rm{ram}}=P_{\rm{thm}}/P_{\rm{ram}}$, where $P_{\rm{thm}}$ is
the thermal pressure of the gas.  Figure
\ref{fig:beta_ram_beta_mag_2000au_mag_mu8OA} shows the evolution of
the ram beta parameter $\beta_{\rm{ram}}$ outside the outflow
($v_r<0$, outside the sky-blue line) and the plasma beta parameter
$\beta_{\rm{plasma}}=P_{\rm{thm}}/P_{\rm{mag}}$ inside the outflow
($v_r>0$, inside the sky-blue line) at the cut-plane of $y=0$ for
model MF8.

Panel (a) of Figure \ref{fig:beta_ram_beta_mag_2000au_mag_mu8OA} shows
that $\beta_{\rm{ram}}$ (color map outside the outflow) is
asymmetrically distributed and its value in the upper region is
smaller than that in the lower region before the unipolar outflow is
driven.  The results indicate that the asymmetric accretion due to the
turbulence of the initial cloud core generates such different ram
pressure distributions at the initial accretion stage if the
turbulence dominates.  Therefore, driving outflow is initially delayed
and suppressed in the region with large values of the ram pressure,
and the outflow is driven earlier in the region with a smaller value
of the ram pressure.

Figure \ref{fig:beta_ram_beta_mag_2000au_mag_mu8OA} shows that the
low-$\beta_{\rm{ram}}$ region gradually expands to
$\sim10^3\thinspace\rm{au}$ in the upper region from panels (b) to
(f), indicating that the ram pressure by the infalling envelopes
increases and suppresses the outflow driving as time proceeds.  Figure
\ref{fig:beta_ram_beta_mag_2000au_mag_mu8OA} also shows that
low-$\beta_{\rm{plasma}}$ region expands inside the region of the
driven unipolar outflow.  Therefore, it is expected that the ram
pressure $P_{\rm{ram}}$ keeps overcoming the magnetic pressure
$P_{\rm{mag}}$ in the further subsequent evolution and no outflow
driving is sustained in the upper low-$\beta_{\rm{ram}}$ region.


\subsection{Protostellar rocket effect}
\label{subsec:rocket_effect}

\begin{figure*}
  \begin{center}
    \includegraphics[clip,width=160mm]{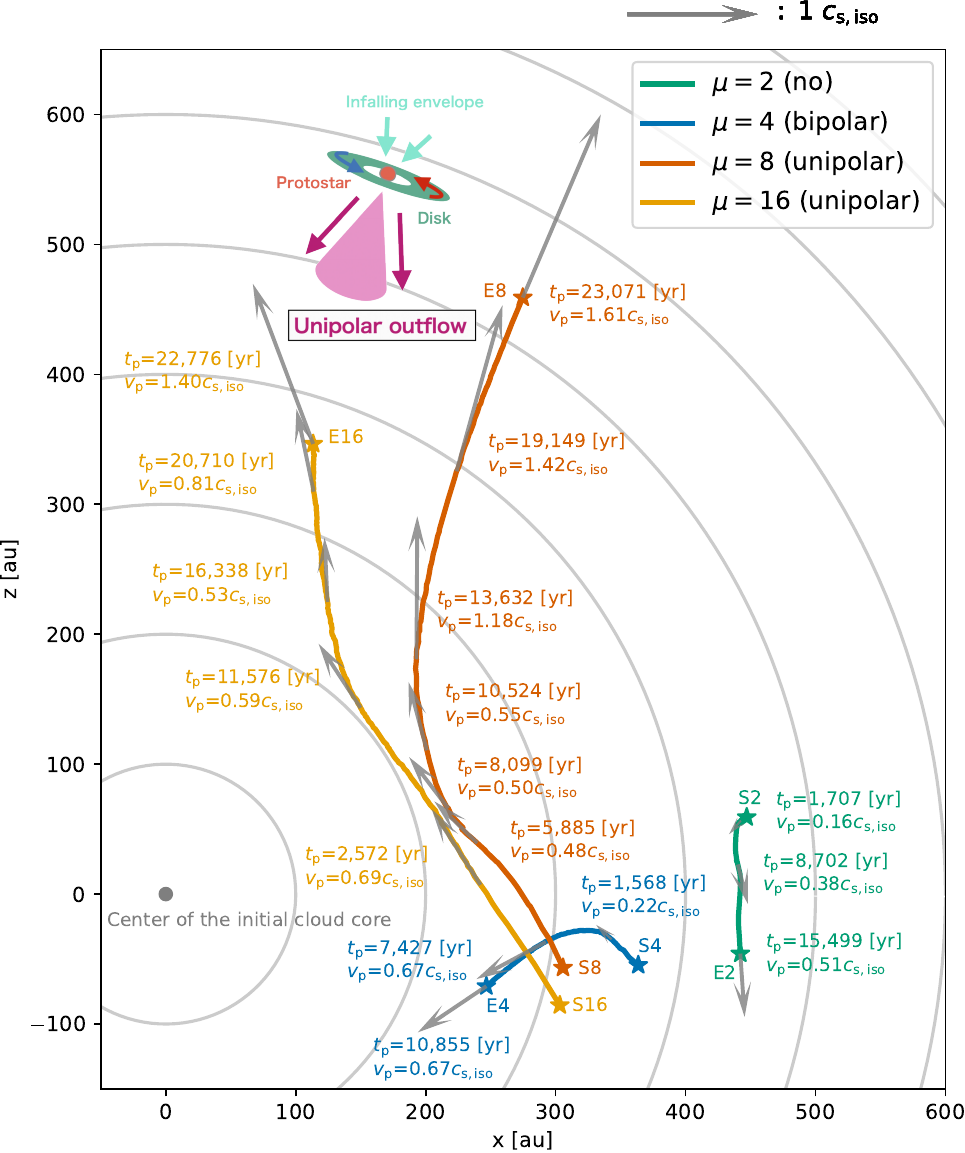}
  \end{center}
  \caption{Projected protostellar trajectories on the $x-z$ plane
    relative to the center of the initial cloud core (gray point) for
    all models.  $t_{\rm{p}}$ is the elapsed time after the protostar
    formation.  $v_{\rm{p}}$ shows the velocity of the protostar.
    Gray arrows indicate the projected protostellar velocity on the
    $x-z$ plane.  The reference arrow plotted on the top right
    corresponds to $1\thinspace c_{\rm{s,iso}}=1.9\times10^4\cms$.
    The symbols ``S'' and ``E'' denote the positions of the protostar
    at its formation ($t_{\rm{p}}=0$) and at the end of the
    simulation.  The numbers beside the symbols ``S'' and ``E'' mean
    the values of the dimensionless mass-to-flux ratio $\magmu$.  A
    schematic illustration on the \textit{protostellar rocket} induced
    by the unipolar outflow is plotted near the positions of ``E8''
    and ``E16''.  Gray circles indicate the same distances from the
    center of the initial cloud core.
    \label{fig:trajectory_xz_mag_mu8OA}
  }
\end{figure*}

Figure \ref{fig:trajectory_xz_mag_mu8OA} shows the projected
protostellar trajectories on the $x-z$ plane relative to the center of
the initial cloud core for all models.  As shown in Figure
\ref{fig:trajectory_xz_mag_mu8OA}, the protostellar systems driving
the unipolar outflow in models MF8 and MF16 move from the inner to the
outer regions of their parent cloud cores at a distance of
approximately $5\times10^2\thinspace\rm{au}$ in the timescale of
$t_{\rm{p}}\sim2\times10^4\thinspace\rm{yr}$ because the unipolar
outflow ejects the protostellar system due to the linear momentum
transport from the unipolar outflow to the protostellar system.
Figure \ref{fig:trajectory_xz_mag_mu8OA} also shows that the projected
protostellar velocities with the unipolar outflow highly increase as
time proceeds, suggesting that the protostellar systems are
accelerated by driving the unipolar outflow.  This phenomenon is
similar to the launch and propulsion of a rocket.  In the following,
we refer to the protostellar systems with the acceleration by the
linear momentum transport of the driven unipolar outflow as the
\textit{protostellar rocket}.

The protostellar rocket amplifies the relative velocity of the
infalling envelopes towards the central protostar and protoplanetary
disk.  Therefore, the results suggest that the increase of the ram
pressure by the infalling envelopes, in the upper region as shown in
panels (b) to (f) in Figures \ref{fig:ratio_PramPmag_2000au_mag_mu8OA}
and \ref{fig:beta_ram_beta_mag_2000au_mag_mu8OA}, is caused by the
combination with such the protostellar rocket and the infalling
envelopes.

One interesting finding is that the driving additional new outflows to
the different directions against the unipolar outflow are prevented
once the protostellar rocket forms, suggesting that the unipolar
outflow is sustained via the above phenomenon.  Furthermore, the
protostellar rocket feedbacks itself to evolve and enhance the
unipolar outflow driving further as shown in Figure
\ref{fig:surface_density_2000au_mag_mu8OA}, resulting that the
feedback system performs to be the instability state.  Hereafter, we
refer to the instability state as the \textit{protostellar rocket
  effect}.

We note that the protostellar systems driving no outflow and the
bipolar outflow also move shorter distances of
$\sim10^2\thinspace\rm{au}$ comparing those driving the unipolar
outflows due to the linear momentum transport from initial chaotic
accretions by the turbulence.  The velocities of the protostellar
systems driving the bipolar outflow and no outflow remain subsonic
throughout the simulations.


\subsection{observational signature of unipolar outflows}
\label{subsec:obs_sign_unip}

\begin{figure*}
  \begin{center}
    \includegraphics[clip,width=155mm]{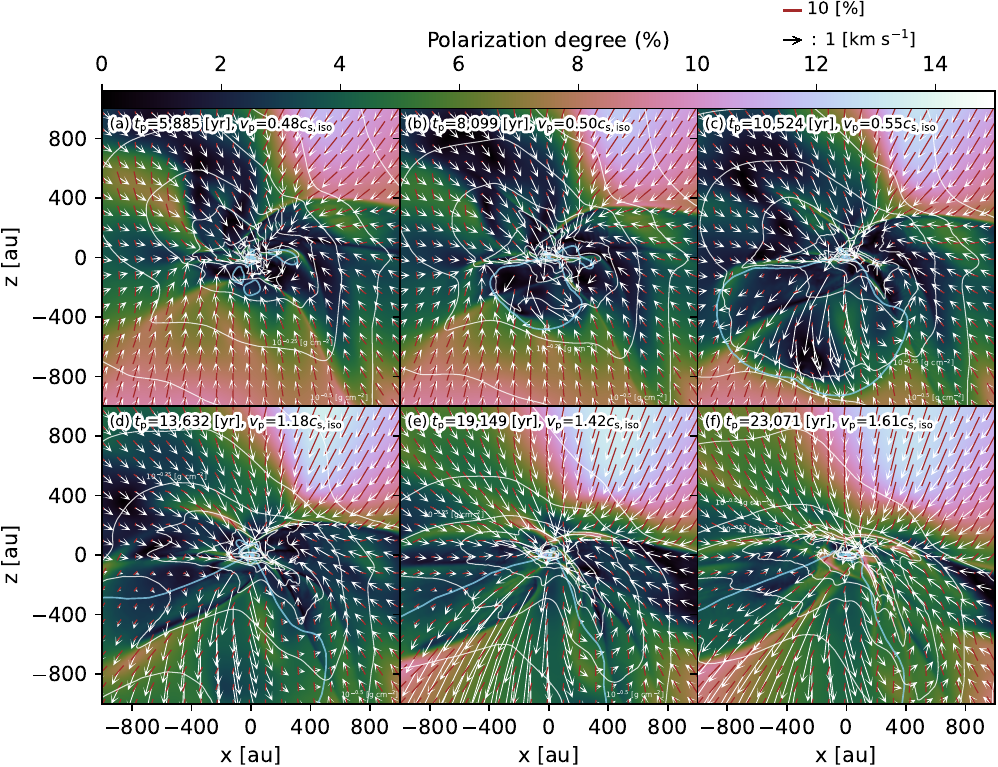}
  \end{center}
  \caption{Expected linear polarization along the $y$-direction for
    the model of the unipolar outflow (model MF8).  The color maps in
    each panel show the polarization degree.  The polarization degree
    vectors are plotted by red-brown bars in each panel.  The elapsed
    times $t_{\rm{p}}$ of each panel are the same as in Figure
    \ref{fig:surface_density_2000au_mag_mu8OA}.  White lines show the
    contours of the surface density in Figure
    \ref{fig:surface_density_2000au_mag_mu8OA} ranging from $-0.75$ to
    $1.25$ in $0.25$ steps on a logarithmic scale.  White arrows of
    the panels show the cut-plane velocity at $y=0$.  The references
    of the polarization degree vector and cut-plane velocity are
    plotted on the top right.  Sky-blue solid lines show the velocity
    contours at $v_r=0$, tracing the front lines of the outflowing
    gas.  The origin of a coordinate system is shifted to the center
    of mass of the system.
    \label{fig:pol_map_mag_mu8OA}
  }
\end{figure*}

Although the unipolar outflows have been detected in many
observations, it is possible that the bipolar outflow is observed as
the unipolar outflow due to extinction effects by the geometry and/or
the surrounding gas.  We propose the expected linear polarization maps
of the thermal emission from dust grains aligned with the magnetic
field as the observational signature to identify whether the unipolar
outflow is real or not.

Figure \ref{fig:pol_map_mag_mu8OA} shows the expected polarization
maps of the thermal emission from dust grains aligned with the
magnetic field for model MF8 in which the unipolar outflow is driven.
The polarization maps are calculated from the relative Stokes
parameters $q$ and $u$ by using the method described in
\citet{2011PASJ...63..147T} (see also Appendix \ref{app:polarization}
for details).  Panel (a) of Figure \ref{fig:pol_map_mag_mu8OA} shows
that the expected polarization degree in both upper and lower regions
has a large value of $\sim10\thinspace\%$ at the scale of
$\sim10^3\thinspace\rm{au}$.  The result suggests that the expected
polarization in the upper region is roughly the same as that in the
lower region before driving the unipolar outflow.  However, panels (b)
to (c) of Figure \ref{fig:pol_map_mag_mu8OA} show that the expected
polarization degree inside the region of the driven unipolar outflow
(inside the sky-blue line) considerably decreases to
$\sim0-5\thinspace\%$.  As shown and pointed out in
\citet{2011PASJ...63..147T}, the protostellar outflow has a low
polarization degree because the toroidal components cancel out each
other inside the outflow region.  The depolarization along the line of
sight therefore emerges by the configuration of the twisted magnetic
field around the unipolar outflow.

During the subsequent evolution from panels (d) to (f) of Figure
\ref{fig:pol_map_mag_mu8OA}, the expected polarization in the upper
region gradually increases to $\sim14\thinspace\%$ and has a large
value relative to that in the lower region whereas the polarization
values becomes slightly large inside the unipolar outflow.  The
results indicate that the expected polarization is quite different
between the lower and upper regions once after driving the unipolar
outflow.  Thus, the unipolar outflow can be identified by the
observation of the thermal dust polarization, although the detectable
sensitivity should be investigated in our subsequent work.  It should
be pointed out in Figure \ref{fig:pol_map_mag_mu8OA} that the
accretion streamers caused by the turbulent accretions of the
infalling envelope have a low expected polarization degree.


\section{Summary and Discussion}
\label{sec:summary_and_discussion}

Recent observations on star-forming regions have revealed that the
protostellar outflows driven by the YSOs exhibit a variety of
asymmetrical features, in particular the unipolar outflows.  Although
the observations have reported the unipolar outflows, the formation
and early evolution of the unipolar outflows driven by low-mass
protostars are yet unclear.  This study investigates the formation and
early evolution of the protostellar outflows with the asymmetrical
features by using the three-dimensional non-ideal MHD simulations of
the gravitational collapse of magnetized turbulent isolated low-mass
cloud cores.  This paper presents, for the first time to our
knowledge, the formation of the unipolar outflow in the early
evolution of the low-mass protostellar system.  Our results and
findings are summarized as follows.

\begin{enumerate}
\item[1.]  The unipolar outflows are driven by the protostellar
  systems formed in the weakly magnetized cloud cores with the
  dimensionless mass-to-flux ratios of $\magmu=8$ and $16$.  In
  contrast, the bending bipolar outflow is driven by the protostellar
  system formed in the moderately magnetized cloud core with
  $\magmu=4$.  Furthermore, no outflow is driven by the protostellar
  system formed in the strongly magnetized cloud core with $\magmu=2$.
  The results explain the observed asymmetrical features of the
  protostellar outflows.
\item[2.]  The protostellar system is ejected by the unipolar outflow
  from the central dense region to the outer region of the parent
  cloud core.  As a result, the protostellar system driving the
  unipolar outflow is gradually accelerated as time passes.  The
  protostellar velocity evolves to be from subsonic to supersonic by
  the acceleration.  This is very similar to the launch and propulsion
  of a rocket, and so we call the protostellar system ejected and
  accelerated by the unipolar outflow the \textit{protostellar
    rocket}.
\item[3.]  We find that the subsequent additional new outflows cannot
  be driven by the protostellar rocket until the end of the
  simulation.  This is because they are suppressed by the ram pressure
  of the infalling envelopes, which is enhanced by the acceleration of
  the protostellar rocket itself.  In the context of low-mass star
  formation, the unipolar outflows can set the protostellar rocket to
  be the instability state, and we call this phenomenon the
  \textit{protostellar rocket effect}.
\end{enumerate}

The remaining question is how the outflow morphologies change with the
different turbulent energies $E_{\rm{turb}}$ from the one adopted in
this study.  Our results show that the unipolar outflows are driven
with $E_{\rm{mag}}/E_{\rm{turb}}<1$, while the bipolar outflow is
driven with $E_{\rm{mag}}/E_{\rm{turb}}\sim1$.  This implies that the
outflow morphologies may depend on the ratio of the magnetic and
turbulent energies of the parent cloud core,
$E_{\rm{mag}}/E_{\rm{turb}}$.

We suggest that the outflow morphologies can be used as a new tracer
to indirectly estimate the magnetic field strengths of the parent
molecular cloud cores.  The measurements of the magnetic field
strengths using the Zeeman effect show that the observed typical cloud
cores are slightly supercritical with the dimensionless mass-to-flux
ratios of $\mu_{\rm{obs}}\sim2$ \citep[e.g.,][]{2008ApJ...680..457T,
  2008A&A...487..247F, 2012ARA&A..50...29C}, which means that the
magnetic pressure is not sufficient to prevent the gravitational
collapse of the cloud cores.  The measurements using the
Davis-Chandrasekhar-Fermi (DCF) method also suggest that the cloud
cores are magnetized with $\mu_{\rm{obs}}\sim2-3$
\citep[e.g.,][]{2006MNRAS.369.1445K, 2023ApJ...952...29K}.  The cloud
cores may actually form from somewhat subcritical initial conditions
\citep[e.g.,][]{2017ApJ...846..122P, 2020ApJ...900..181K,
  2021MNRAS.504.2381Y, 2022MNRAS.515.5689P, 2022ApJ...941..122C,
  2023ApJ...952...29K}.  However, the measurements of $\mu_{\rm{obs}}$
suffer fundamentally from their statistical and systematic
uncertainties such as the selection bias towards sources with strong
magnetic fields for the Zeeman effect observations and the
overestimates of the magnetic field strengths in the DCF method
\citep{2021ApJ...919...79L}.

In our simulations, a bipolar outflow forms in a cloud core with
$\magmu=4$ ($\magmu_{\rm{const}}=2$) and ${\cal{M}}_{\rm{s}}=0.86$.
Typical cloud cores have the turbulence of
${\cal{M}}_{\rm{s}}\lesssim1$ \citep[e.g.,][]{2007prpl.conf...33W}.
Therefore, our results suggest that the formation of bipolar outflows
in turbulent cloud cores requires the relatively strong magnetic field
of the parent cloud cores.  In contrast, the unipolar morphology may
form with the relatively weak magnetic field of the parent cloud
cores.  \citet{2004A&A...426..503W} show that bipolar outflows are
observed more frequently than unipolar outflows.  This suggests that
the typical molecular cloud cores tend to have relatively strong
magnetic fields.  Note, however, that the turbulence strength is fixed
in our current study and the impact of different turbulence strengths
remains unclear.  We will investigate the relation between the outflow
morphologies, magnetic field strengths ($E_{\rm{mag}}$), and
turbulence strengths ($E_{\rm{turb}}$) in our future work.

The protostellar rocket effect would drive shock waves into the
envelope around the protostar, which may possibly be detected by some
chemical shock tracers in the molecular line emissions.  However, the
detectability of the shock waves depends on the chemical species of
the shock tracers and their lifetimes in the post-shock waves.  In
addition, even just accreting gas, without the protostellar rocket
effect, is capable of driving the accretion shocks.  Therefore, the
detectability of the shock waves in the ambient cloud material needs
to be investigated in detail to identify the probable chemical shock
tracers with their characteristics.

This study ignores the Hall effect due to the computational cost to
include it.  Here, we briefly discuss possible impact of the Hall
effect on our conclusion.  The magnetic braking is strengthened or
weakened by the Hall effect when the rotation and magnetic field
vectors are aligned or anti-aligned, and the Hall effect introduces
some interesting phenomena in the formation and early evolution of the
protostars and protoplanetary disks in collapsing cloud cores
\citep[e.g.,][]{1999MNRAS.303..239W, 2011ApJ...733...54K,
  2011ApJ...738..180L, 2012MNRAS.422..261B, 2015ApJ...810L..26T,
  2016MNRAS.457.1037W, 2017PASJ...69...95T, 2018MNRAS.475.1859W,
  2018MNRAS.480.4434W, 2021MNRAS.507.2354W}.  If the Hall effect is
strong enough even in our simulation environments, we speculate that
when the rotation and magnetic field vectors are aligned, the magnetic
field may be twisted more strongly and the outflow is stronger, making
it harder for the unipolar outflow to form.  On the other hand, when
the rotation and magnetic field vectors are anti-aligned, the magnetic
field may be twisted more weakly and the outflow is less likely to
overcome the ram pressure, which may lead to the formation of the
unipolar outflow.  However, it should be noted that the magnetic
resistivity depends on the dust model, cosmic ray ionization rate, and
also on the strength of the magnetic field
\citep[e.g.,][]{1999MNRAS.303..239W, 2002ApJ...573..199N,
  2014A&A...571A..33P, 2016A&A...592A..18M, 2018MNRAS.476.2063W,
  2019MNRAS.484.2119K, 2022ApJ...934...88T, 2022MNRAS.515.2072K,
  2023MNRAS.521.2661K, 2023ASPC..534..317T}.  These predictions will
be verified by future simulations including the Hall effect.

How long the protostellar rocket effect continues in subsequent
evolution is still an open question.  More evolved protostellar
rockets are expected to be distributed outside the region of the
initial cloud core as long as the unipolar outflow is maintained.
Thus, the unipolar outflow and the protostellar rocket may be
observationally identified by combining the proper motions, the launch
direction of the unipolar outflow on the plane of the sky, and the
morphologies of the infalling envelope cavity using, for example, the
Global Astrometric Interferometer for Astrophysics (GAIA), the James
Webb Space Telescope (JWST), ALMA, and/or other methods.  However,
even if peculiar proper motions are detected, it may be difficult to
distinguish between protostellar rockets being its origin and, for
example, stellar encounters being its origin.  The long-term evolution
and observability of protostellar rockets will be considered in our
future studies.


\section*{Acknowledgments}  
We thank Dr. Tomoaki Matsumoto, Dr. Kazuya Saigo, Dr. Konstantin
Batygin, and Dr. Keiichi Wada for the comments and fruitful
discussions.  We are also grateful to the anonymous referee for the
fruitful suggestions and insightful comments that helped to improve
the manuscript.  Numerical computations were carried out on Cray
XC50 at Center for Computational Astrophysics, National Astronomical
Observatory of Japan.  D.T. acknowledges the fellowship by the Japan
Society for the Promotion of Science (JSPS).  This work is supported
by JSPS KAKENHI Grant Numbers JP21J23102 (D.T.), JP21H00048 (S.T.),
JP21H04495 (S.T.), JP18H01247 (Y.S.) and JP23H01212 (Y.S.).  Y.T. is
supported by JST FOREST Grant Number JPMJFR2234.  S.T. is supported
by NAOJ ALMA Scientific Research Grant Number 2022-20A.

\vspace{5mm}
\software{
  Matplotlib \citep{2007CSE.....9...90H},
  Mayavi \citep{2011CSE....13b..40R},
  Numpy \citep{2020Natur.585..357H}
}

\appendix



\section{Calculation of the polarization of the thermal emission from aligned dust grains}
\label{app:polarization}

The linear polarization of the thermal emission from dust grains
aligned with the magnetic field reveals the configuration of the
magnetic field around the driven outflows.  This appendix describes
the calculation of the polarization of the thermal emission from
aligned dust grains for Figure \ref{fig:pol_map_mag_mu8OA}.

The linear polarization of the thermal emission from aligned dust
grains is calculated from the Stokes parameters $Q$ and $U$.  As
written in \citet{2011PASJ...63..147T} (we also refer the reader to
\citet{1985ApJ...290..211L, 1990ApJ...362..120W, 2000ApJ...544..830F}
for more details), the relative Stokes parameters $q$ and $u$ are
substituted for $Q$ and $U$ by assuming the grain properties and
temperature are constant in the entire region of the calculation and
neglecting the background starlight absorption and the dust
scattering:
\begin{align}
  q &= \int \rho \cos 2\psi \cos^2\gamma \thinspace ds, \label{eq:rel_st_q} \\
  u &= \int \rho \sin 2\psi \cos^2\gamma \thinspace ds, \label{eq:rel_st_u}
\end{align}
where $\psi$ is the angle between the north and the projected magnetic
field on the plane of the sky, and $\gamma$ is the angle between the
plane of the sky and the local direction of the magnetic field.  The
integrations $\int ds$ are conducted along the line of sight normal
vector $\bm{n}$ in a range from $-2R_{\rm{c}}$ to $2R_{\rm{c}}$
centered on the protostar.  The vertical and horizontal axes of the
observational grid are chosen along the unit vectors
$\bm{e}_{\eta}=(\bm{e}_{z}-(\bm{e}_{z}\cdot\bm{n})\bm{n})/
|\bm{e}_{z}-(\bm{e}_{z}\cdot\bm{n})\bm{n}|$ and
$\bm{e}_{\xi}=\bm{e}_{\eta}\times\bm{n}$, where $\bm{e}_{z}$ is the
unit vector in the direction of the $z$-axis.  The line of sight
direction is chosen as $\bm{n}=-\bm{e}_{y}$ in this study, resulting
in $\bm{e}_{\eta}=\bm{e}_{z}$ and $\bm{e}_{\xi}=\bm{e}_{x}$, where
$\bm{e}_{x}$ and $\bm{e}_{y}$ are the unit vectors in the direction of
the $x$ and $y$ axes.  From the relative Stokes parameters $q$ and
$u$, the polarization angle $\chi$ and the polarization degree vector
$\pvec$ on the observational grid are calculated as
\begin{align}
  \chi  &= \left\{
  \begin{array}{ll}
    \frac{1}{2} \arctan\frac{u}{q} & \rm{for}~{\it{\cos\chi}} \neq 0 \\
    \frac{\pi}{2}                  & \rm{for}~{\it{\cos\chi}} =    0
  \end{array}
  \right.,
  \label{eq:pangle} \\
  \pvec &\equiv
  \begin{pmatrix}
    p_{\xi} \\
    p_{\eta}
  \end{pmatrix}
  =
  \begin{pmatrix}
    p_{\rm{pol}} \sin \chi \\
    p_{\rm{pol}} \cos \chi
  \end{pmatrix},
  \label{eq:pvec}
\end{align}
where $\cos\chi$ and $\sin\chi$ are determined by
\begin{align}
  \cos \chi  &= \sqrt{\frac{\cos2\chi+1}{2}}, \\
  \sin \chi  &= \left\{
  \begin{array}{ll}
    \frac{\sin 2\chi}{2\cos\chi}  & \rm{for}~{\it{\cos\chi}}\neq0 \\
    1                             & \rm{for}~{\it{\cos\chi}}=0
  \end{array}
  \right., \\
  \cos 2\chi &= \frac{q}{\sqrt{q^2+u^2}}, \\
  \sin 2\chi &= \frac{u}{\sqrt{q^2+u^2}}.
  \label{eq:chi_angles}
\end{align}
$p_{\rm{pol}}$ is the ratio of the polarized to total intensities
calculated as
\begin{align}
  p_{\rm{pol}} &= p_{\rm{0}} \frac{\sqrt{q^2+u^2}}{\Sigma-p_{\rm{0}}\Sigma_{2}},
  \label{eq:ppol} \\
  \Sigma       &= \int \rho ds,
  \label{eq:sigma} \\
  \Sigma_{2}   &= \int \rho \left(\frac{\cos^2\gamma}{2}-\frac{1}{3}\right) ds,
  \label{eq:sigma2}
\end{align}
where the parameter $p_{\rm{0}}$ is chosen as $p_{\rm{0}}=0.15$ to fit
the highest degree of polarization observed in typical interstellar
clouds.  We note that the grain properties and temperature are assumed
to be constant, following the approach most frequently used in
previous studies.  In our next study, observable polarization features
of the unipolar outflows will be investigated in more detail by the
Stokes parameters taking into account the self-consistent calculation
of the grain alignment efficiencies and the wavelength-dependent full
radiative transfer calculation \citep[e.g.,][]{2017A&A...603A..71R,
  2019MNRAS.482.2697S}.


\bibliography{export-bibtex, export-bibtex-2, takaishids_article-bibtex}  
\bibliographystyle{aasjournal}


\end{document}